\documentclass[aps,preprint]{revtex4}
\usepackage{epsfig}
\usepackage{amsmath}

\begin{document}
\title
{\bf Non-unitarity and non-reciprocity in scattering from real potentials in presence of confined non-linearity}    
\author{Zafar Ahmed}
\email{zahmed@barc.gov.in}
\affiliation{Nuclear Physics Division, Bhabha Atomic Research Centre,Mumbai 400 085, India}
\date{\today}

\begin{abstract}
Investigations  of scattering in presence of non-linearity which have just begun require the confinement of both the potential, $V(x)$, and the non-linearity, $\gamma f(|\psi|)$. 
There could be two options for the confinement. One is the finite support on $x \in [-L,L]$ and the other one is on $x \in [0,L]$. 
Here, we consider real Hermitian potentials and report a surprising disparate behaviour of these two types of confinements.
We prove that in the first option  the symmetric potential enjoys reciprocity of both reflectivity ($R$) and transmitivity ($T$) and their unitarity. More interestingly, the asymmetry in $V(x)$ causes non-unitarity ($ R+T\ne 1$) and the non-reciprocity (reciprocity) of $T (R)$. On the other hand, the second option  of confinement gives rise to an essential non-unitarity even when  $V(x)$ is symmetric about a point in $[0,L]$. In the absence of symmetry there occurs non-reciprocity of both $R$ and $T$.

\end{abstract}
\maketitle
We  are generally taken by surprise when we first learn that the probability of quantal reflection, $R$,
and transmission, $T$, do not depend on whether the particles are incident from left or right side of a
real Hermitian potential. This is called reciprocity  of reflectivity, $R$, and transmitivity $T$. Also $R$ and $T$ display unitarity by adding to 1, for all energies of incidence. Therefore these two are features 
of scattering from real Hermitian potentials involving
linear Schro{\"o}dinger equation (SE).

Here we find that the non-linear Schr{\"o}dinger (NSE) equation for  real Hermitian potentials presents
interesting surprises. We show that non-unitarity and non-reciprocity in scattering can occur even for  real
Hermitian potentials involving NSE in presence of confined non-linearity.

The present work is motivated to study the scattering involving non-linear Schr{\"o}dinger equation (NSE)[1-5] 
\begin{equation}
-{\partial^2 \Psi \over \partial x^2}+V(x)\Psi+\gamma f(|\Psi|) \Psi=-i{\partial \Psi \over \partial t}
\end{equation}
with confined non-linearity which has just begun [1-3]. In NSE (1), when $f(|\Psi|)=|\Psi|^2$, this is the well known cubic Schr{\"o}dinger equation which is derived from Duffing-Lorentz model of linear response of the matter to the electromagnetic waves, by neglecting the imaginary (dissipation) term. Due to the inherent similarity of the Schr{\"o}dinger and the Helmontz wave equation, in optics this quadratic term $(|\Psi|^2)$ is also known as Kerr's non-linearity.

Despite the non-linearity the NSE (1) can be checked to follow the  continuity equation 
\begin{equation}
{d \over dt} <\Psi^*| \Psi>
= {i\hbar \over 2m} 
\left [\Psi^* {d\Psi \over dx}- \Psi {d\Psi^* \over dx} \right]_{x_1} ^{x_2}.
\end{equation}
which is similar to that of SE. This allows us to calculate the current density in the usual way. Let us remark that, this simplicity could be deceptive to think for example that if the non-linearity is  confined spatially, the NSE for real Hermitian $V(x)$ will again enjoy the similar features of scattering namely, reciprocity and unitarity as that of SE. The present works wards off this presumption.

Recently, the scattering from complex (optical) PT-symmetric potentials  in SE has received renewal of interest and thrown several novel phenomena [7,9] like spectral singularity, time reversed lasers (anti-lasers), lasing with anti-lasing, invisibility and transparency. Further scattering from a complex potential involving confined non-linearity has been investigated [1-3] to study the role of non-linearity on the spectral-singularity in complex PT-symmetric cases. Due to intensive focus [1-5] on complex potentials, we feel that the question of the features of scattering from a real Hermitian potential involving NSE remains  unaddressed.

When the scattering potential in SE is optical (complex) the reflection and transmission are again reciprocal but without unitarity, if the potential is spatially symmetric. If not then $R$ displays non-reciprocity [6,7]. Therefore, the reflectivity for the PT-symmetric complex potentials which have imaginary part as anti-symmetric displays non-reciprocity.

However, complex PT-symmetric Scarf II potential in SE in some parametric regime is known [8] to enjoy both unitarity and reciprocity of reflection despite non-Hermiticity and PT-symmetry.

In this work we consider both $V(x)$ and $f(|\psi|)$ as real  in NSE (1). Using $\Psi=\psi(x) e^{-iEt/\hbar}, k=\sqrt{E}$ with $2m=1=\hbar$ in NSE (1) one gets
\begin{subequations}
\label{allequations} 
\begin{eqnarray}
&&{d^2\psi(x) \over dx^2}+k^2 V(x) \psi(x) + \gamma f(|\psi(x)|)\psi(x)=0, \quad   
x \in D_1\label{equationa}
\\
&&{d^2\psi(x) \over dx^2}+k^2 \psi_(x)=0,\quad x \in D_2,
\label{equationb}
\end{eqnarray}
\end{subequations}
For scattering both $V(x)$ and $f|\psi|$ need to be  confined in space. There are two options for doing this, the first one is 
\begin{equation}
D_1=[-L,L], \quad  \mbox{and} \quad D_2 = (-\infty, -L) \cup (L, \infty).
\end{equation}
The second one is
\begin{equation}
D_1=[0,L], \quad \mbox{and} \quad D_2 = (-\infty, 0) \cup (L, \infty).
\end{equation}
In Ref. [3,5] the first option is chosen whereas in Ref. [1,2] the second option has been chosen however without a justification. Here we study both and find their surprising disparate behaviour, in that the symmetric $V(x)$ in the first option of confinement leads to unitarity, whereas a symmetric potential in $[0,L]$ gives rise to an essential non-unitarity in scattering.

In this work, we take both $V(x)$ and $f(|\psi|)$ as strictly real to bring out the claimed features of scattering arising from NSE (3) in two types of confinements (4,5). The solution of (3) for the incidence from left:
\begin{eqnarray}
&&\psi(x<-L)= A_l e^{ikx} + B_l e^{-ikx} \\ \nonumber
&&\psi(|x|\le L)=\alpha~ u(x)+ \beta~ v(x)\\ \nonumber
&&\psi(x \ge L)=C_l e^{ikx},
\end{eqnarray}
and for the  incidence from right are:\\
\begin{eqnarray}
&&\psi(x>L)= A_r e^{ikx} + B_r e^{-ikx} \\ \nonumber
&&\psi(|x|\le L)= \gamma~ u(x) + \delta~ v(x) \\ \nonumber
&&\psi(x \ge L)=C_r e^{ikx}.
\end{eqnarray}
By matching  these solutions (6,7) and their derivative
at $x=\pm L$ and introducing the convenient notations:
$u_1=u(L), v_1=v(L), u_2=u(-L), v_2=v(-L)$,
we get the reflection and transmission amplitudes

\begin{eqnarray}
r_{left}= 
\frac{[u^\prime_2 v^\prime_1-u^\prime_1 v^\prime_2]+ik[u^\prime_1 v_2+u_1 v^\prime_2]
-ik[u_2 v^\prime_1+u^\prime_2 v_1]+k^2[u_1 v_2-u_2 v_1]}
{[u^\prime_2 v^\prime_1-u^\prime_1 v^\prime_2]+ik[u^\prime_1 v_2-u_1 v^\prime_2]
-ik[u_2 v^\prime_1-u^\prime_2 v_1]+k^2[u_1 v_2-u_2v_1]}
\end{eqnarray}
\begin{eqnarray}
r_{right}=
\frac{[u^\prime_2 v^\prime_1-u^\prime_1 v^\prime_2]+ik[u^\prime_2 v_1+u_2 v^\prime_1]
-ik[u_1 v^\prime_2+u^\prime_1 v_2]+k^2[u_1 v_2-u_2v_1]}
{[u^\prime_2 v^\prime_1-u^\prime_1 v^\prime_2]+ik[u^\prime_1 v_2-u_1 v^\prime_2]
-ik[u_2 v^\prime_1-u^\prime_2 v_1]+k^2[u_1 v_2-u_2v_1]}
\end{eqnarray}
\begin{eqnarray}
t_{left}=
\frac{-2ik[u_2v^\prime_2-u^\prime_2v_2]}{[u^\prime_2 v^\prime_1-u^\prime_1 v^\prime_2]+ik[u^\prime_1 v_2-u_1 v^\prime_2]
-ik[u_2 v^\prime_1-u^\prime_2 v_1]+k^2[u_1 v_2-u_2v_1]}
\end{eqnarray}
\begin{eqnarray}
t_{right}=
\frac{-2ik[u_1v^\prime_1-u^\prime_1 v_1]}{[u^\prime_2 v^\prime_1-u^\prime_1 v^\prime_2]+ik[u^\prime_1 v_2-u_1 v^\prime_2]
-ik[u_2 v^\prime_1-u^\prime_2 v_1]+k^2[u_1 v_2-u_2v_1]}
\end{eqnarray}

Here $u(x)$ and $v(x)$ are two real linearly independent solutions of the NSE (3). For numerical computations
we take $u(0)=1, u^\prime(0)=0; v(0)=0, v^\prime(0)=1$ as initial values so the wronskian at $x=0$ is
\begin{equation}
W(0)=[u(0) v^\prime (0)- u^\prime (0) v(0)]=1.
\end{equation}
We then  integrate the NSE (3) step by step towards
both $x=\pm L$. The values of these functions at the end points are used in Eqs. (8-11) to determine the $R(E)=|r|^2$ and $T(E)=|t|^2$.   

Notice that the all the terms in the square brackets in the numerators and denominators of Eqs. (8-11) are real by virtue of the reality of both $V(x)$ and $f(|\psi|)$. The denominators of these
equations are identical. The imaginary terms in the numerators of $r_{left}$ and $r_{right}$ (8,9) have only the signs opposite. The crucial consequence of this is that
\begin{equation}
r_{left} \ne r_{right}\quad \mbox{but} \quad R_{left}=R_{right}.
\end{equation}
This proves the common feature of reciprocity of reflectivity for both SE and NSE (3) for real $V(x)$ and $f(|\psi|)$. It does not matter whether real potential is spatially symmetric or non-symmetric. 

The wronskians in the numerators of
both $t_{left}$ and $t_{right}$ are constant of scattering process and do not change from its  initial value of 1 (12) for SE [$\gamma=0$, in Eq. (3)]. Most interestingly for NSE ($ \gamma \ne 0$) the wronskians 
\begin{equation}
W_1=[u_1v^\prime_1-u^\prime_1 v_1] \quad \mbox{and} \quad W_2=[u_2v^\prime_2-u^\prime_2v_2]
\end{equation}
are not independent of $x$. Instead these are interesting function  of energy and the parameters of the potential when calculated at end points $x=\pm L$ as can be seen by working out
\begin{equation}
{d\over dx} [u(x)v^\prime(x)-v(x) u^\prime(x)] = {dW \over dx}==-\gamma [f(|u|)-f(|v|)] u(x) v(x).
\end{equation}
When $V(x)$ and $f(|\psi|)$ are symmetric function of $x$, we have $u(x)$ and $v(x)$ of definite parity: even and odd, respectively as
\begin{equation}
u_1=u_2,~ v_1=-v_2,~ u^\prime_1=-u^\prime_2,~ v^\prime_1=v^\prime_2.
\end{equation}
Consequent to this, $W_1=W_2$, implying the reciprocity of transmitivity
\begin{equation}
t_{left}=t_{right}\quad {and} \quad T_{left}= T_{right}.
\end{equation}
Eventually, the unitarity occurs which can be checked by using the conditions (16) in Eqs. (8-11).
\begin{equation}
R={(k^2u_1 v_1+u_1^\prime v_1^\prime)^2
\over (k^2 u_1 v_1 - u_1^\prime v_1^\prime)^2 + k^2(u_1v_1^\prime+u_1^\prime v_1)^2}, \quad T={k^2(u_1v_1^\prime-u_1^\prime v_1)^2 \over 
(k^2 u_1 v_1 - u_1^\prime v_1^\prime)^2 + k^2(u_1v_1^\prime+u_1^\prime v_1)^2}
\end{equation}
add to 1, for both the left and the right incidence.
Most interestingly when $V(x)$ is non-symmetric Eq.(16) does not hold giving rise to $W_1\ne W_2$ and hence the transmitivity becomes non-reciprocal
\begin{equation}
T_{left} \ne T_{right}
\end{equation}
and non-unitarity 
\begin{equation}
R+T_{left} \ne 1 \quad \mbox{and} \quad R+T_{right} \ne 1.
\end{equation}
takes place even without the potential being non-Hermitian.

Now let us follow the second option of confining
both $V(x)$ and $f(|\psi|)$ in the domain $[0,L]$.
So the solution for the NSE (2) for the left incidence
can be written as
\begin{eqnarray}
&&\psi(x<0)=A e^{ikx}+ B e^{-ikx},\\ \nonumber
&&\psi(0<x \le L)= C u(x) + D v(x),\\ \nonumber
&&\psi(x>L)=F e^{ikx}.
\end{eqnarray}
By matching  these solutions at $x=0,L$, we get the reflection and transmission amplitudes
\begin{equation}
r_{left}={k^2v+u^\prime+ik(v^\prime-u) \over k^2 v-u^\prime+ik(v^\prime+u)} \quad t_{left}={2ik(u v^\prime-u^\prime v) \over k^2 v-u^\prime+ik(v^\prime+u)}.
\end{equation}
Here  $u,v,u',v'$ written without any argument mean the values of these functions at the end point ($x=L$). We integrate (3) using the same initial values  as mentioned above Eq. (12) from $x=0$ to $x=L$ and use the end values of the various functions in Eq. (22) to evaluate $R$ and $T$ numerically.

These formula are convenient to show the non-unitarity in case of NSE (3). Using $W=uv'-u'v$ in (22), we find that
\begin{equation}
R+T=\frac{k^4u'^2+k^2v'^2+k^2u^2+2k^2W[2W-1]}{k^4u'^2+k^2v'^2+k^2u^2+2k^2W},
\end{equation} 
which equals 1 only if $W=1$. This happens in the case of SE. For NSE as discussed above $W \ne 1$, instead it becomes function of energy. Thus we
have non-unitarity despite real Hermitian potential.

We present our calculations to demonstrate various features of scattering involving NSE. Here, we use  two profiles for $f(|\psi|)$
one is saturating type: $f_S(|\psi|)={1 \over 1+|\psi|^2}$ as suggested in [3] and the other one is Kerr's non-linearity: $f_K(|\psi|) = |\psi|^2$. For the potential we use the Gaussian profile $ V_G(x)=V_0 e^{-x^2}$. 
In all the calculations we use $2m=1=\hbar^2$. The length parameter of confinement as $L=5$ and the strength of non-linearity $\gamma=1$. In all the Figs. 1-5, we plot $R$, $T$
and $R+T$ in parts a,b,c. The thick/blue curves present the incidence from left and the thin/red 
curves present the incidence from the right. In the cases of invariance with respect to the side of incidence these pairs of curves would coincide with each other to appear just one to demonstrate the said invariance. 

Figs. 1-3 are for the first option of confinement: $x\in [-L,L]$ (4). Fig. 1 confirms the reciprocity of $R$ and $T$ and unitarity of scattering when  $V(x)$ is symmetric. Non-linearity used here is $f_S$, however, we have checked that the use of $f_K$ (non-saturating) does not show any qualitatively disparate behaviour.

Fig. 2 displays the reciprocity of $R$ and non-reciprocity of $T$ along with the non-unitarity of scattering when $V(x)$ is non-symmetric. Fig. 3 displays the same effect when $f_K$  is used. We, however, find that $f_K$ non-linearity is stronger which changes the even trend of the variation of $R$ and $T$ as compared to that of $f_S$ in Fig. 2.

For the second option (5) of confinement in $[0,L]$,  we choose $V_{\mu}(x)=V_0 \exp[-(x-\mu L)^2]$.
Fig. 4 confirms  the reciprocity of both $R$ and $T$ and non-unitarity as $V_{\mu}(x) (\mu=1/2)$ is symmetric about $x=L/2$ in $[0,L]$. If $V(x)$ does not have a point of symmetry in $[0,L]$, notice the  non-reciprocity of both $R$ and $T$ in Fig. 5 (a,b). 
We have studied various other potentials  (attractive, and  repulsive) for $V(x)$ with several saturating and non-saturating profiles for $f(|\psi|)$ to verify the qualitative robustness of our various results presented here.

Finally, we would like to remark that various features of scattering from real Hermitian potentials  involving non-linear Schr{\"o}dinger equations proved  and demonstrated here are new and robust. We   
conclude that the non-linear Schr{\"o}dinger equation can give rise to both non-reciprocity and non-untarity in scattering despite the potential being real Hermitian. The disparate  behaviour of non-linear Schr{\"o}dinger equation for two kinds of confinements presented here  is thought provoking. We hope
that these results will be helpful in both experimental
and theoretical investigations of scattering in presence of confined non-linearity which have just begun recently.
\begin{figure}
\centering
\includegraphics[width=4 cm,height=4 cm]{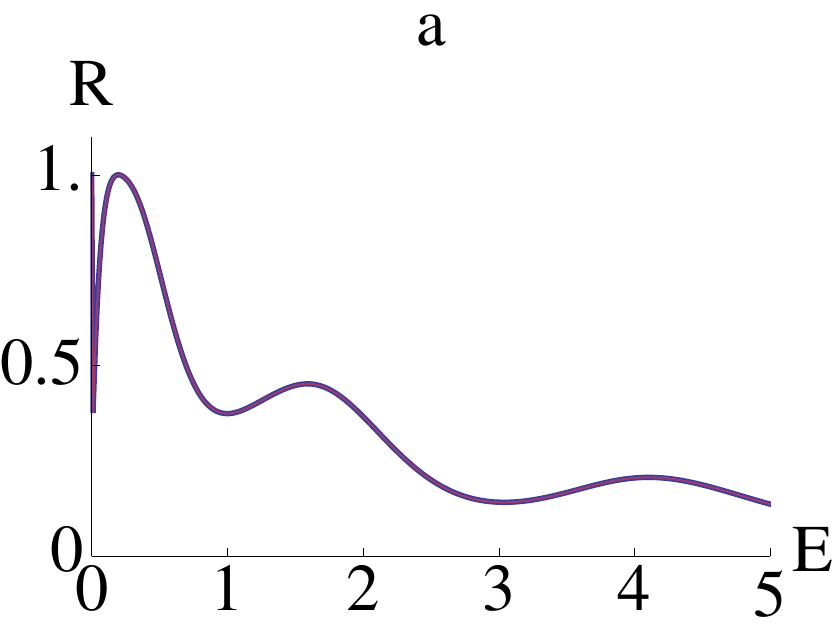}
\hskip .5 cm
\includegraphics[width=4 cm,height=4 cm]{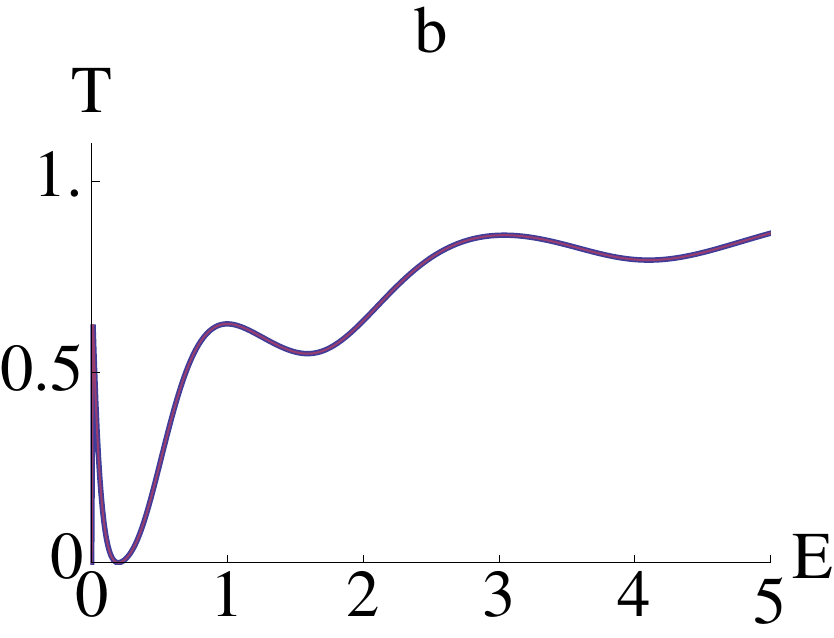}
\hskip .5 cm
\includegraphics[width=4 cm,height=4 cm]{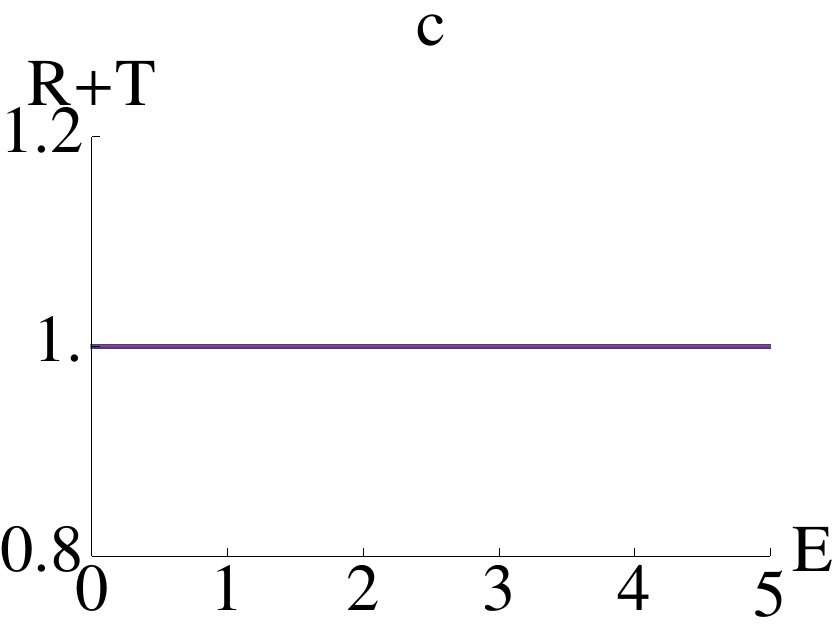}
\caption{The plots of $R,T$ and $R+T$ for $V_G(x)$ (symmetric) and saturating non-linearity $f_S$. $V_0=-3$,  $\gamma=1$ and $L= 5$. The  reciprocity of both $R$ and $T$ and unitarity is observed as the curves for left (thick/blue) and right (thin/red) incidence have merged in all three parts. The confinement bot $V(x)$ and $f_S(|\psi|)$ is on $[-L,L]$  }
\end{figure}
\begin{figure}
\centering
\includegraphics[width=4 cm,height=4 cm]{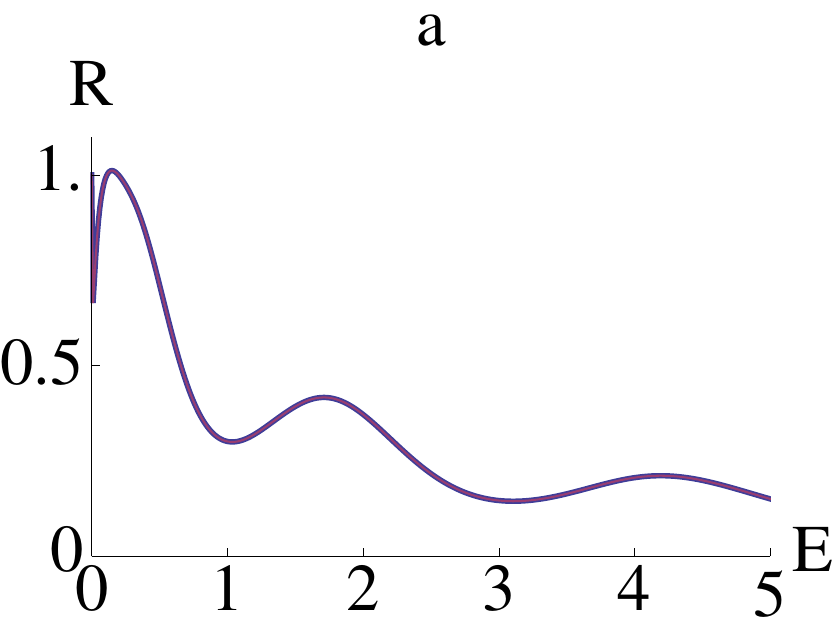}
\hskip .5 cm
\includegraphics[width=4 cm,height=4 cm]{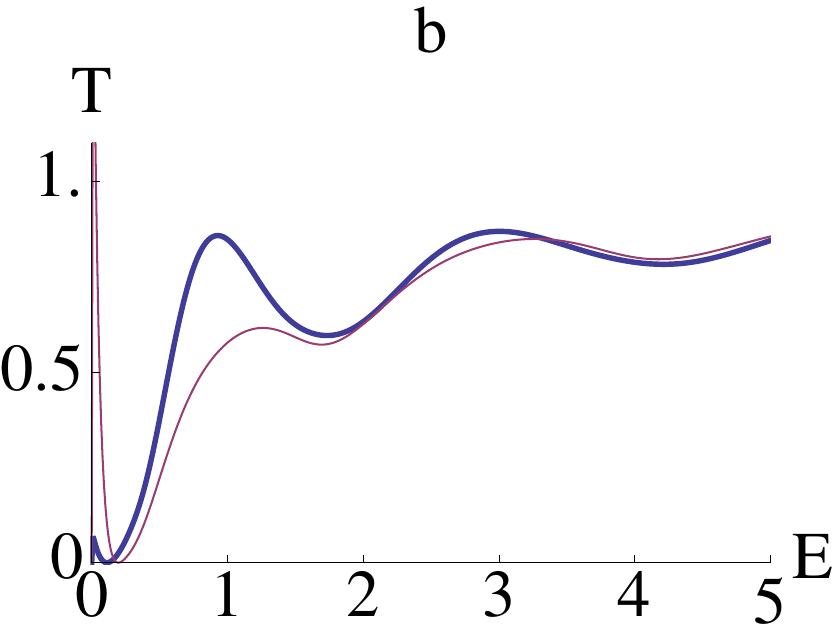}
\hskip .5 cm
\includegraphics[width=4 cm,height=4 cm]{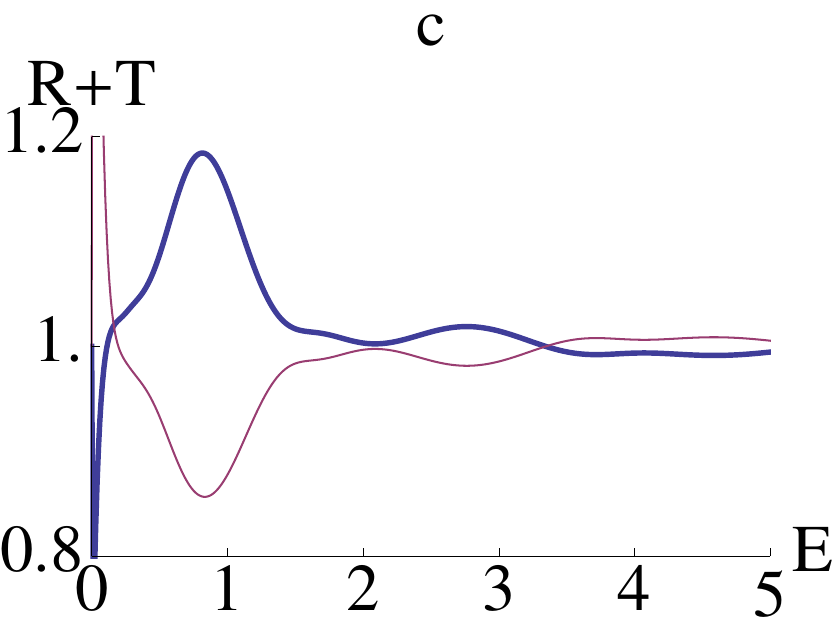}
\caption{The same as in Fig. 1, but the potential is asymmetric: $V(x\le 0)=-3 ~e^{-x^2}, V(x>0)= -3 ~e^{-3x^2/2}$. The effect of asymmetry can be seen as the $T_{left}$ and $T_{right}$  curves deviate but $R_{left}$ and $R_{right}$ are coincident. In c the non-unitarity of scattering is is displayed as $R+T$ for both the left  and the right  incidence deviate from 1}
\end{figure}
\begin{figure}
\centering
\includegraphics[width=4 cm,height=4 cm]{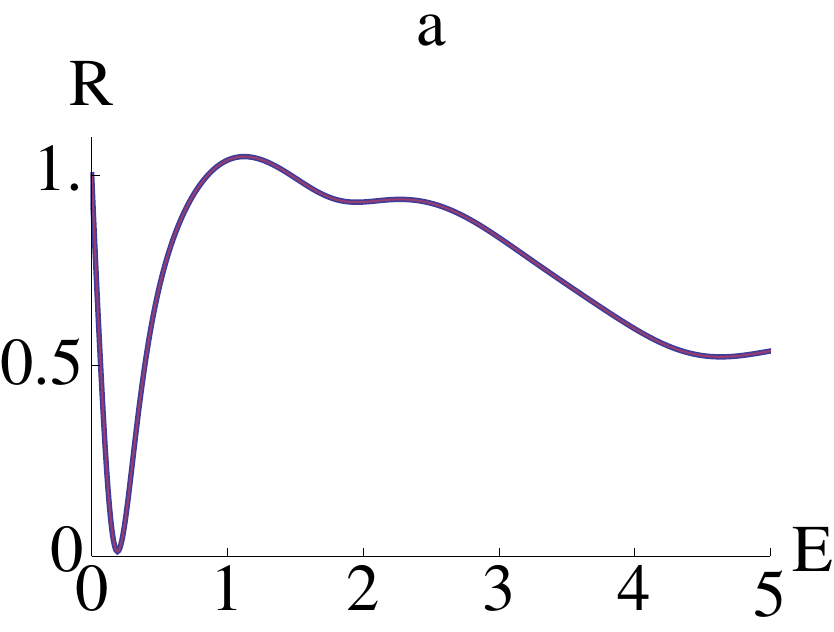}
\hskip .5 cm
\includegraphics[width=4 cm,height=4 cm]{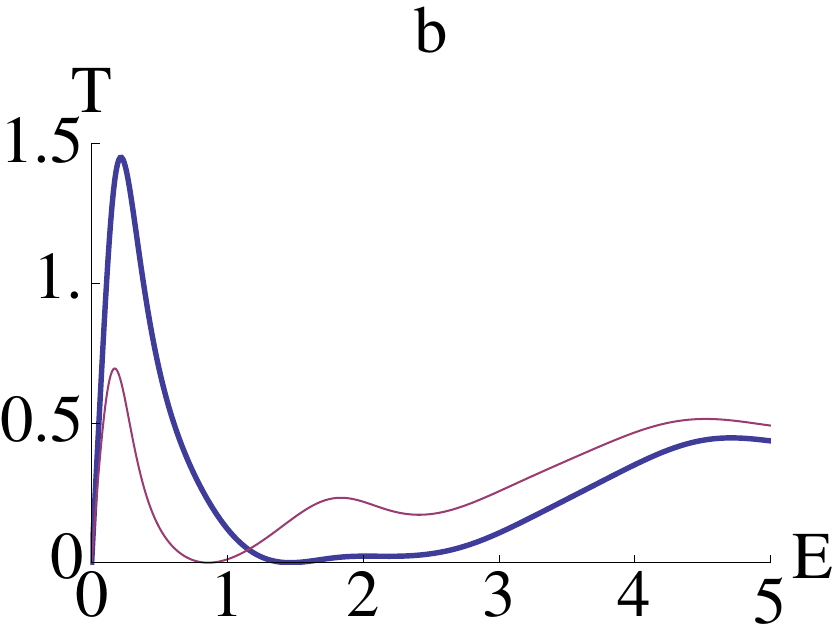}
\hskip .5 cm
\includegraphics[width=4 cm,height=4 cm]{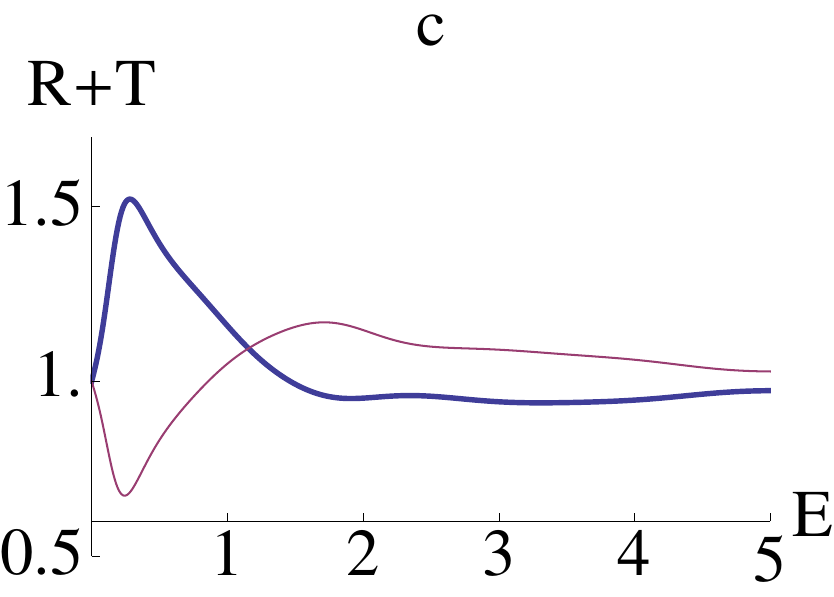}
\caption{The same as in Fig. 2, excepting that that the
non-linearity is Kerr's type: $f_K(\psi|)=|\psi|^2$.  It shows that Kerr's non-saturating non-linearity is a stronger one to change even the trend of variation of $R$ and $T$ for the same fixed potential, compare with Fig. 2.}
\end{figure}
\begin{figure}
\centering
\includegraphics[width=4 cm,height=4 cm]{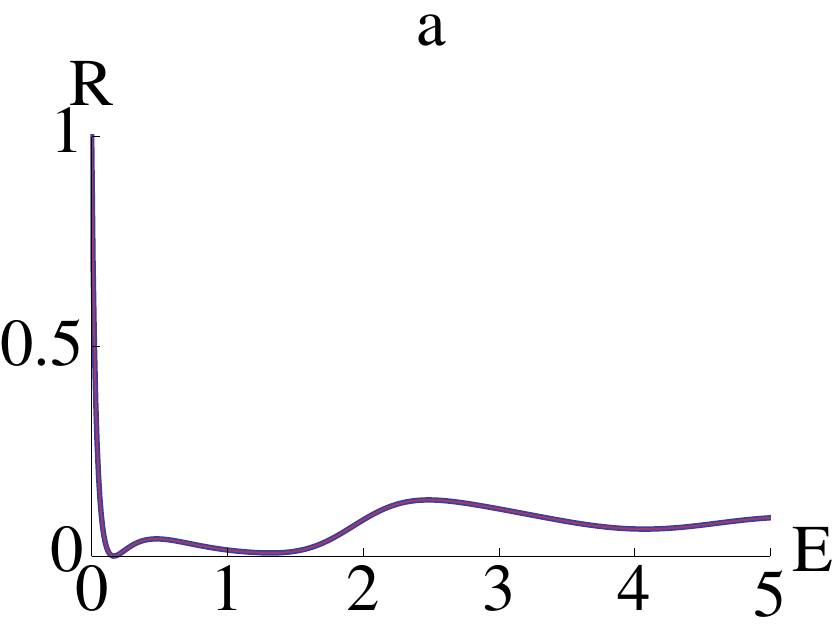}
\hskip .5 cm
\includegraphics[width=4 cm,height=4 cm]{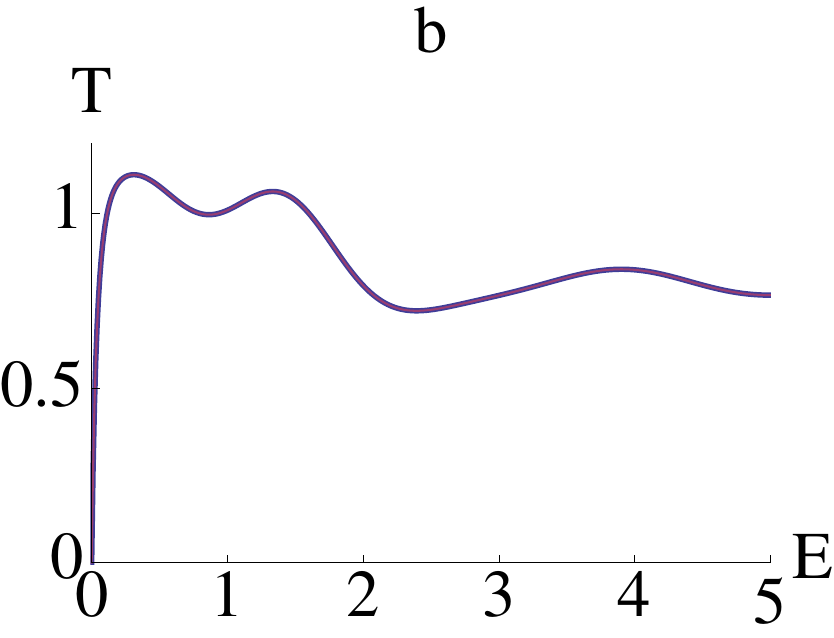}
\hskip .5 cm
\includegraphics[width=4 cm,height=4 cm]{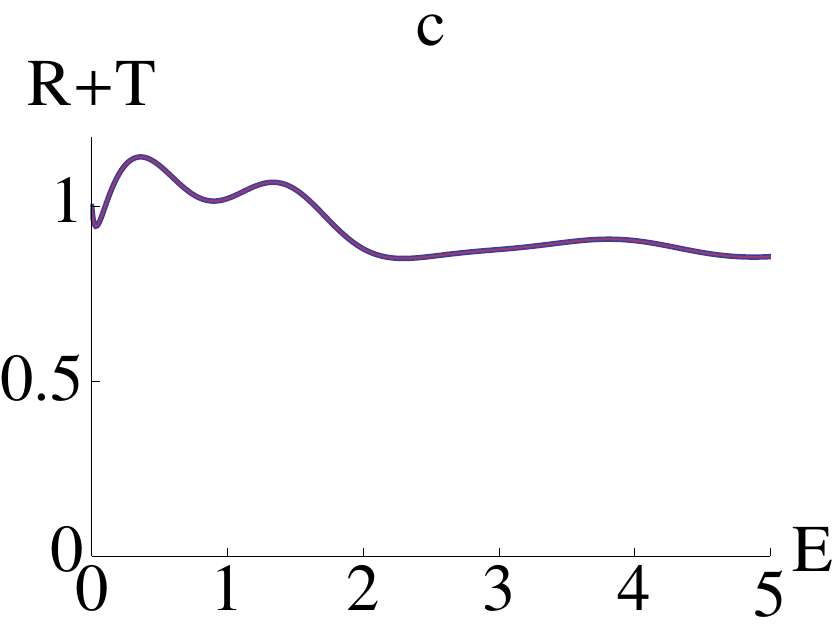}
\caption{The same as in Fig. 1 for the confinement on $[0,L]$. Here we use $V_\mu$ and $f_{K}$ (see the text) $\mu=.5,\gamma=1,V_0=3,L=5$. Like Fig.1 blue/thick and red/thin curves merge to show insensitivity to the left and right incidence. But unlike Fig. 1, here non-unitarity takes place as $R+T \ne 1$}
\end{figure}
\begin{figure}
\centering
\includegraphics[width=4 cm,height=4 cm]{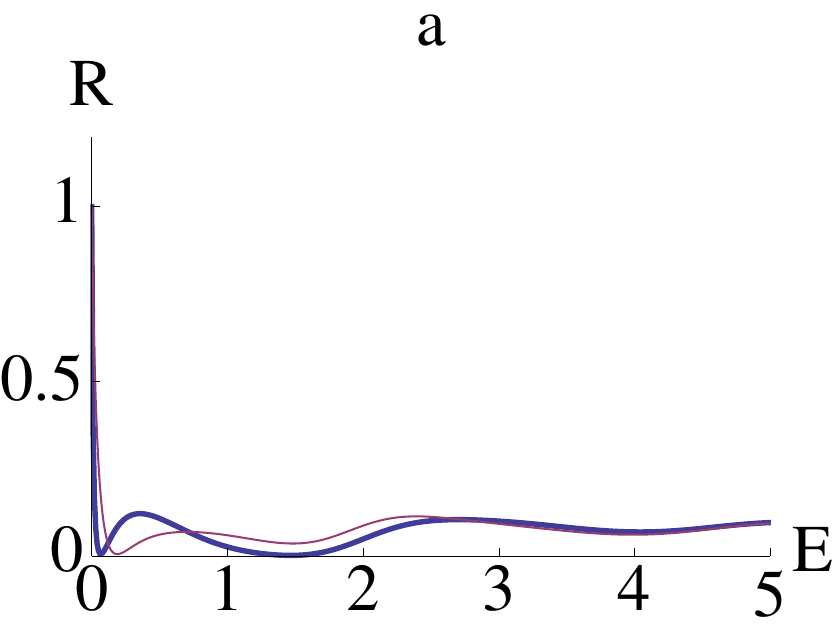}
\hskip .5 cm
\includegraphics[width=4 cm,height=4 cm]{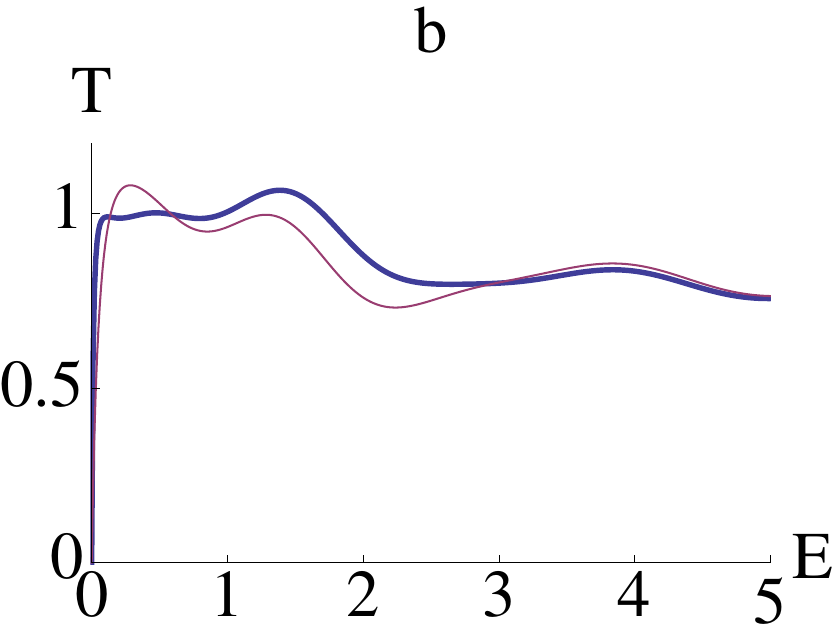}
\hskip .5 cm
\includegraphics[width=4 cm,height=4 cm]{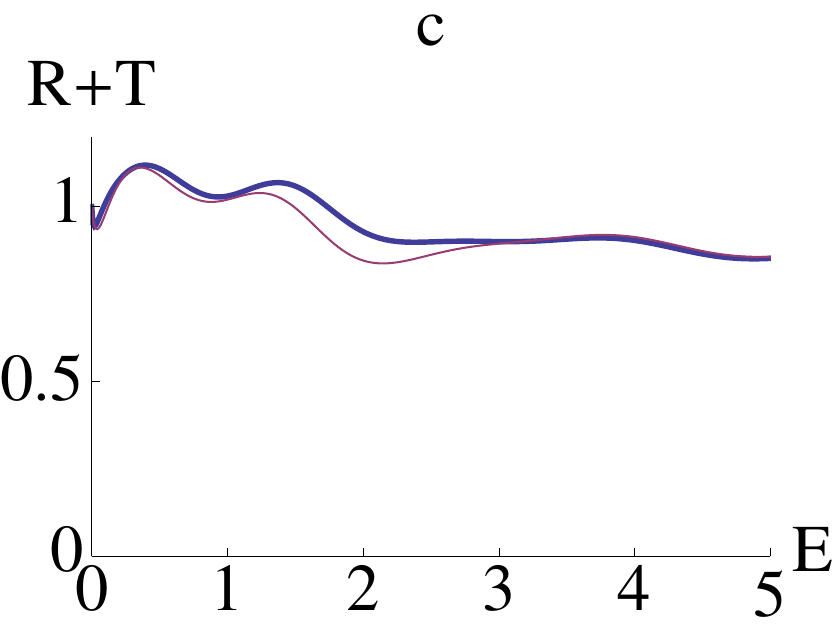}
\caption{The same as Fig. 4, but $\mu=.4$. $V_{\mu}(x)$ is non-symmetric in $[0,L].$ Notice the non-unitarity and non-reciprocity of both $R$ and $T$ as the blue/thick and thin/red curves deviate from each other.}
\end{figure}

\end{document}